\documentclass[reprint,notitlepage,superscriptaddress,preprintnumbers,nofootinbib,nobibnotes,amsmath,amssymb,aps,twocolumn]{revtex4-1}
\usepackage{graphicx,grffile}
\usepackage{dcolumn}
\usepackage{bm}
\usepackage{hyperref}
\usepackage{aas_macros,braket}

\makeatletter
\def\lst@lettertrue{\let\lst@ifletter\iffalse}
\makeatother

\begin{document}

\title{Constraining graviton non-Gaussianity through the CMB bispectra}

\author{Valerio De Luca} 
\affiliation{D\'epartement de Physique Th\'eorique and Centre for Astroparticle Physics (CAP), Universit\'e de Gen\`eve, 24 quai E. Ansermet, CH-1211 Geneva, Switzerland}
\author{Gabriele Franciolini} 
\affiliation{D\'epartement de Physique Th\'eorique and Centre for Astroparticle Physics (CAP), Universit\'e de Gen\`eve, 24 quai E. Ansermet, CH-1211 Geneva, Switzerland}
\author{Alex Kehagias} 
\affiliation{Physics Division, National Technical University of Athens, 15780 Zografou Campus, Athens, Greece}
\author{Antonio Riotto}
\affiliation{D\'epartement de Physique Th\'eorique and Centre for Astroparticle Physics (CAP), Universit\'e de Gen\`eve, 24 quai E. Ansermet, CH-1211 Geneva, Switzerland}
\affiliation{CERN, Theoretical Physics Department, Geneva, Switzerland}
\author{Maresuke Shiraishi}
\affiliation{Department of General Education, National Institute of Technology, Kagawa College, 355 Chokushi-cho, Takamatsu, Kagawa 761-8058, Japan}

\date{\today}


\begin{abstract}

  Tensor non-Gaussianities are a key ingredient to test the symmetries and the presence of higher spin fields during the inflationary epoch. Indeed, the shape of the three point correlator of the graviton is totally fixed by the symmetries of the de Sitter stage and, in the case of parity conservation, gets contributions only from the ordinary gravity action plus a higher derivative term called the  (Weyl)$^3$ action. We discuss current and future bounds on the three point tensor contribution from the  (Weyl)$^3$ term using cosmic microwave background (CMB) bispectra. Our results indicate that forthcoming experiments,  such as LiteBIRD, CMB-S4 and CORE, will detect the presence of the  (Weyl)$^3$ term if $M_p^4 L^4 \sim 10^{17} r^{-4}$, where $L$ parametrizes the strength of the  (Weyl)$^3$ term and $r$ is the tensor-to-scalar ratio, which corresponds to $L\gtrsim  3.2 \times 10^5 M_p^{-1}$,  while the current upper limit is $M_p^4 L^4 = (1.1 \pm 4.0) \times 10^{19} r^{-4}$~(68\%\,CL).

\end{abstract}

\maketitle


\section{Introduction}

Current observational evidences suggest that the universe experienced a period of accelerated expansion in its early stages which goes under the name of cosmological inflation \cite{lr}. In this  phase the spacetime geometry is approximately described by the de Sitter (dS) metric. One of its main unknown features is its particle content, which one could be able to specify looking at signatures in the present day observables with current/future experiments. 
Additionally, since inflation could have taken place at energies much higher than the ones reachable by collider experiments, it provides a unique opportunity to test high energy physics \cite{cc1,cc2,cc3,cc4,Bartolo:2017szm}.

One very important property of inflation is that, by being described by an approximately de Sitter metric, it enjoys all its background symmetries. As we shall see, those symmetries are so powerful that allow to predict the properties of many observables. During this period, the spacetime isometries form an (approximate) SO(4,1) group which can be identified with the conformal group of a CFT$_3$ acting on perturbations extending on scales larger than the Hubble radius. One of the signatures of the inflationary epoch that could have reached us is given by the gravitational waves and, in particular, the graviton non-Gaussianities (NGs), which are dictated by symmetry arguments. Thus, any observational confirmation of the graviton NG would pave the way to a deeper understanding of the theory of gravity and particle content during inflation. Indeed, symmetry arguments (better to say the isometries of de Sitter) constrain the three point correlator (dubbed sometime the bispectrum) of the graviton to be  conformal invariant. Its shape is fixed (when parity is preserved) by two interacting terms, the ordinary Einstein-Hilbert gravity action and the higher-order derivative action called the  (Weyl)$^3$ term. 

Since a sizable contribution of the (Weyl)$^3$ term has been related to possible causality violation \cite{Camanho:2014apa}, which are supposedly cured by the presence of a tower of higher-spin states, an experimental confirmation of a large higher derivative tensor NG would suggest the presence of an infinite tower of higher spin (HS) particles during inflation. Indeed, one particularly interesting setup where the power of symmetries manifests itself is in theories with higher spin fields. The possible signature of the presence of massive (or partially massless) fields during inflation has been extensively studied both theoretically \cite{hsm1,hsm2,hsm3,hsm4,hsm6,hsm7,hsm10,hsm11,hsm12} and observationally \cite{hsm5,hsm8,hsm9}. In fact, in the massless case,  there are very strong theorems constraining the existence of  HS fields in Minkowski spacetime (see no-go theorems \cite{nogo1,nogo2,nogo3}) which can be avoided in curved spacetime settings \cite{yesgo}. A noticeable example is provided by the Vasiliev theory, which describes the dynamics of interacting massless higher spin particles in curved spacetime \cite{vas1,vas2,vas3} and admits solutions which are asymptotically de Sitter (or anti de Sitter) \cite{vas4,vas5,Aros:2017ror}. In general, the theory contains an infinite tower of interacting massless gauge fields of any spin and, since it accommodates a massless spin-2 particle, the graviton, it is therefore an appealing theory of gravity.

The bispectrum of cosmic microwave background (CMB) is one of the cleanest observables of graviton NGs. So far, signatures of various NG shapes and their detectabilities have been theoretically examined (e.g. Refs.~\cite{Shiraishi:2010kd,Shiraishi:2011st,Shiraishi:2012sn,Shiraishi:2013vha,Shiraishi:2013kxa,Shiraishi:2016yun,Meerburg:2016ecv,Domenech:2017kno,Tahara:2017wud,Bartolo:2018elp,Kothari:2019yyw}). Moreover, some of these shapes have already been tested with the real CMB data \cite{Shiraishi:2013wua,Shiraishi:2014ila,Ade:2015ava,Shiraishi:2017yrq,Akrami:2019izv}. No detection of $> 2\sigma$ signal yields useful constraints on the background theories (see Ref.~\cite{Shiraishi:2019yux} for a review).

This paper aims at constraining the graviton NG induced by (Weyl)$^3$ term by employing the CMB bispectra. CMB signatures of the same NG template have already been analyzed in an early work \cite{Shiraishi:2011st}, while how much they are constrained from current and forthcoming CMB observations remains undiscussed.%
\footnote{
  In Ref.~\cite{Shiraishi:2011st}, a graviton NG from parity-violating  (Weyl)$^3$ cubic term $f(\phi) \widetilde{W} W^2$ has also been studied. This has already been constrained using the CMB temperature bispectrum in Ref.~\cite{Shiraishi:2014ila}.}
We here address these issues for the first time. 

Through the temperature data analysis, we find no evidence of the graviton NG. This provides an upper bound on $M_p^4 L^4$ as in Eq.~\eqref{eq:WMAP_limit}.  On the other hand, our forecasted analysis shows that its minimal detectable size shrinks by some orders of magnitudes by taking B-mode polarization into account. This encourages follow-up investigations in next decadal B-mode experiments such as LiteBIRD \cite{Hazumi:2012aa}, CMB-S4 \cite{Abazajian:2016yjj}, and CORE \cite{Delabrouille:2017rct}.

The paper is organized as follows. In Sec.~II we review the tensor NG shapes as dictated by the dS symmetries and discuss the particular example of the Vasiliev higher spin gravity giving rise to such a scenario. In Sec.~III we present the observational test based on the currently available data and the forecasted sensitivity with the next generation of future experiments. The last section is devoted to the conclusions.

\section{Graviton non-Gaussianity in de Sitter spacetime} \label{sec:prim}


During the inflationary epoch, the spacetime is approximately described by a de Sitter metric 
\begin{equation}
	d s^2 = \frac{1}{H^2 \tau^2} \left ( - d \tau^2 + d \vec x^2\right),
\end{equation}
which is invariant under the symmetry group SO(4,1). In the late-time limit ($\tau \to 0$), the background isometries act as a ${\rm CFT}_3$ group containing translations, rotations, dilatations and special conformal transformations. This also goes under the name of dS/CFT$_3$ correspondence. Under such a symmetry, the structure of the two and three point functions is completely fixed and their late-time limit is naturally captured by the correlator structure in the CFT$_3$ dual description. In the case of the graviton three point correlator,  it can be shown that the symmetries dictate the shape of the bispectrum to be induced by  an interacting action \cite{Maldacena:2011nz} composed by the Einstein-Hilbert term
\begin{equation}
S_{\rm EH}=\frac{M_p^2}{2} \int d\tau d^3x \sqrt{-g}
 \left [ R -6 H^2\right],
\end{equation}
 plus an additional (Weyl)$^3$  term as \footnote{One additional possible contribution to the graviton three point function would come from the parity-violating interaction term $\widetilde W W^2$. From now on, we will consider only the  parity-conserving  case. }
\begin{equation}
  S_{W^3}= \frac{M_p^2 L^4}{2} \int d\tau d^3x \sqrt{-g} \,
  W^{ab}{}_{cd} W^{cd}{}_{mn} W^{mn}{}_{ab} . 
   \label{eq:W3_action} 
\end{equation}
The length scale $L$ determines the strength of the  (Weyl)$^3$ term. The explicit expression for the three point correlator is given in Refs.~\cite{Maldacena:2002vr,Maldacena:2011nz}. The same structure is also found working in the three dimensional CFT$_3$ dual description \cite{noianninos,cft1,cft2}.

One explicit example of the theory of gravity giving rise to such a graviton NG is provided by the Vasiliev theory in de Sitter spacetime, where the interactions between the particles are fully dictated by the higher spin gauge group and the spectrum is composed by an infinite tower of massless higher spin fields. The theory depends only on a free parameter identified with the ratio between the reduced Planck mass and the Hubble rate during inflation, $M_p/H$. 
The prediction regarding the tree-level late-time cosmological correlation functions of the graviton in such a theory has been worked out in various works \cite{ann1,ann2,anninos,ann3}, while the full nonlinear result can be computed with the aid of the framework introduced in Ref.~\cite{anninos} making use of the dS/CFT$_3$ correspondence.  In particular, one can analyze the characteristic graviton non-Gaussianity through the three point function predicted by the higher spin gravity and obtained in Ref.~\cite{noianninos}.
The nice property of the higher spin gravity is that the interactions between the particles are completely fixed, thus also the additional scale $L$ is a clear prediction of the theory. In particular, it is given by \cite{noianninos}
\begin{equation}
H L = \left (\frac{8}{270} \right)^{1/4} \simeq 0.4 . \label{eq:hl}
\end{equation}
This means that  in the  Vasiliev theory the higher derivative corrections to the GR graviton three point function are sizable.

Even though the structure of the graviton three point function is robustly fixed by symmetries, different theories of gravity could lead to the prediction of a different scale $L$ which would impact the possibility of a future experimental confirmation through CMB observations. From now on, we will consider $L$ as a free parameter.

\section{Observational tests of tensor non-Gaussianities using the CMB bispectra}
\label{sec:CMB_bis}

Here we examine the detectability of a graviton NG from the  (Weyl)$^3$ cubic action \eqref{eq:W3_action}, as a probe of the higher spin gravity model, using CMB bispectra. Moreover, by constraining the graviton NG with the real CMB data, we test this model. 


\subsection{CMB bispectra}

Since the CMB fields are distributed on the 2 sphere, their magnitudes are quantified through the spherical harmonic decomposition. Tensor-mode harmonic coefficients of temperature and E/B-mode polarization fields ($X = T, E, B$) take the form \cite{Shiraishi:2010sm,Shiraishi:2010kd}
\begin{equation}
a_{\ell m}^{X} = 
(-i)^{\ell} \int \frac{d^3 \vec{k}}{2 \pi^2}
{\cal T}_{\ell}^{X}(k)  \sum_{\lambda = \pm 2} \left(\frac{\lambda}{2}\right)^x \gamma_{\vec{k}}^{(\lambda)} {}_{-\lambda} Y_{\ell m}^*(\hat{k}) , 
\end{equation}
where $x = \begin{cases} 0 & (X = T,E) \\ 1 & (X = B) \end{cases}$, and $\gamma_{\vec{k}}^{(\pm 2)}$ is a helicity-state representation of primordial GW, which is given by 
\begin{equation}
 \gamma_{ij}(\vec{x}) \equiv \int \frac{d^3 \vec{k}}{(2\pi)^3}
 e^{i \vec{k} \cdot \vec{x}} \sum_{\lambda = \pm 2} \gamma_{\vec{k}}^{(\lambda)} e_{ij}^{(\lambda)}(\hat{k}) ,
\end{equation}
with the transverse and traceless polarization tensor $e_{ij}^{(\pm 2)}(\hat{k})$ obeying $e_{ij}^{(\lambda) *}(\hat{k}) = e_{ij}^{(-\lambda)}(\hat{k}) = e_{ij}^{(\lambda)}(- \hat{k})$ and $e_{ij}^{(\lambda)}(\hat{k}) e_{ij}^{(\lambda')}(\hat{k}) = 2 \delta_{\lambda, -\lambda'}$. The linear transfer function ${\cal T}_{\ell}^{X}(k)$ expresses characteristic tensor-mode features, e.g., the integrated Sachs-Wolfe (ISW) amplification for temperature and the reionization bumps for polarization \cite{Pritchard:2004qp}. The CMB bispectrum then takes the form
\begin{eqnarray}
&& \Braket{\prod_{n=1}^3 a_{\ell_n m_n}^{X_n}}
  = \left[ \prod_{n = 1}^3(-i)^{\ell_n} \int_0^\infty \frac{k_n^2 d k_n}{2 \pi^2}
    {\cal T}_{\ell_n}^{X_n}(k_n) \right] \nonumber \\
  \nonumber \\
  && \qquad \times \sum_{\lambda_1, \lambda_2, \lambda_3 = \pm 2} \left(\frac{\lambda_1}{2}\right)^{x_1} \left(\frac{\lambda_2}{2}\right)^{x_2} \left(\frac{\lambda_3}{2}\right)^{x_3} \nonumber \\ 
  && \qquad \times  \left[\prod_{n=1}^3
    \int d^2 \hat{k}_n \, {}_{-\lambda_n} Y_{\ell_n m_n}^*(\hat{k}_n) \right]  \Braket{\prod_{n=1}^3 \gamma_{\vec{k}_n}^{(\lambda_n)}}. \label{eq:CMB_bis_form}
\end{eqnarray}

A helicity-state expression of graviton bispectrum from the  (Weyl)$^3$ cubic action \eqref{eq:W3_action} (and also a formula of induced CMB bispectrum) has already been obtained in Ref.~\cite{Shiraishi:2011st}. Here, we present a mathematically equivalent but alternative form where contractions between unit vectors and polarization tensors are much simplified, reading  
\begin{eqnarray}
  \Braket{\prod_{n=1}^3 \gamma_{\vec{k}_n}^{(\lambda_n)}}
  &=&  (2\pi)^3 \delta^{(3)} \left(\sum_{n=1}^3 \vec{k}_n\right)
f_{k_1 k_2 k_3} \delta_{\lambda_1, \lambda_2} \delta_{\lambda_2, \lambda_3} \nonumber \\ 
&& \times e_{ij}^{(-\lambda_1)} (\hat{k}_1) e_{jk}^{(-\lambda_2)} (\hat{k}_2)
e_{ki}^{(-\lambda_3)} (\hat{k}_3) , \label{eq:h3}
\end{eqnarray}
where
\begin{equation}
f_{k_1 k_2 k_3} \equiv 5760 \left(\frac{H}{M_p}\right)^8 M_p^4 L^4 (k_1 + k_2 + k_3)^{-6} . 
\end{equation}
In a similar way to Ref.~\cite{Shiraishi:2011st}, inserting this into Eq.~\eqref{eq:CMB_bis_form}, we compute angular integrals and helicity summations, and simplify the resulting formula by employing the addition rule of angular momenta. The bottom-line formula has a rotational-invariant shape 
\begin{equation}
\Braket{\prod_{n=1}^3 a_{\ell_n m_n}^{X_n}} 
= \left(
  \begin{array}{ccc}
   \ell_1 & \ell_2 & \ell_3 \\
  m_1 & m_2 & m_3
  \end{array}
  \right) B_{\ell_1 \ell_2 \ell_3}^{X_1 X_2 X_3} ,
\end{equation}
where 
\begin{eqnarray}
  && B_{\ell_1 \ell_2 \ell_3}^{X_1 X_2 X_3}
  = \frac{- ( 8\pi )^{3/2}}{5} \sqrt{\frac{7}{3}}
  (-i)^{\ell_1 + \ell_2 + \ell_3}
  \nonumber \\
  && \qquad \times
  \delta_{x_1 + x_2 + x_3 + \ell_1 + \ell_2 + \ell_3}^{(e)}
  \sum_{L_1 L_2 L_3 } (-1)^{\frac{L_1 + L_2 + L_3}{2}}
  h_{L_1 L_2 L_3}^{0 \ 0 \ 0} \nonumber \\
  && \qquad \times  
  h_{\ell_1 L_1 2}^{2 0 -2}
  h_{\ell_2 L_2 2}^{2 0 -2}
  h_{\ell_3 L_3 2}^{2 0 -2}
  \left\{
  \begin{array}{ccc}
    \ell_1 & \ell_2 & \ell_3 \\
    L_1 & L_2 & L_3 \\
    2 & 2 & 2
  \end{array}
  \right\}
  \int_0^\infty y^2 dy
  \nonumber \\
  && \qquad \times 
  \left[\prod_{n=1}^3 \frac{2}{\pi}  \int_0^\infty k_n^2 dk_n
    {\cal T}_{\ell_n}^{X_n}(k_n) j_{L_n}(k_n y)  \right]
  f_{k_1 k_2 k_3} ,
   \label{eq:CMB_bis}
\end{eqnarray}
with $ \delta_{l}^{(e)} \equiv \begin{cases} 0  & (l = {\rm odd}) \\ 1  & (l = {\rm even}) \end{cases}$ and
\begin{equation}
  h^{s_1 s_2 s_3}_{l_1 l_2 l_3}
\equiv \sqrt{\frac{(2 l_1 + 1)(2 l_2 + 1)(2 l_3 + 1)}{4 \pi}}
\left(
  \begin{array}{ccc}
  l_1 & l_2 & l_3 \\
  s_1 & s_2 & s_3
  \end{array}
 \right) .
\end{equation}
The angle-averaged bispectrum \eqref{eq:CMB_bis} takes a much simpler form than the previous one in Ref.~\cite{Shiraishi:2011st} (although both yield identical result); hence, much more speedy numerical analysis is realized. Note that parity-conserving nature of the graviton bispectrum \eqref{eq:h3} restricts nonvanishing signal to $TTT$, $TTE$, $TEE$, $TBB$, $EEE$, and $EBB$ when $\ell_1 + \ell_2 + \ell_3 = \rm even$, and $TTB$, $TEB$, $EEB$, and $BBB$ when $\ell_1 + \ell_2 + \ell_3 = \rm odd$.

\begin{figure}[t]
  \begin{tabular}{c}
    \begin{minipage}{1.\hsize}
  \begin{center}
    \includegraphics[width = 1.\textwidth]{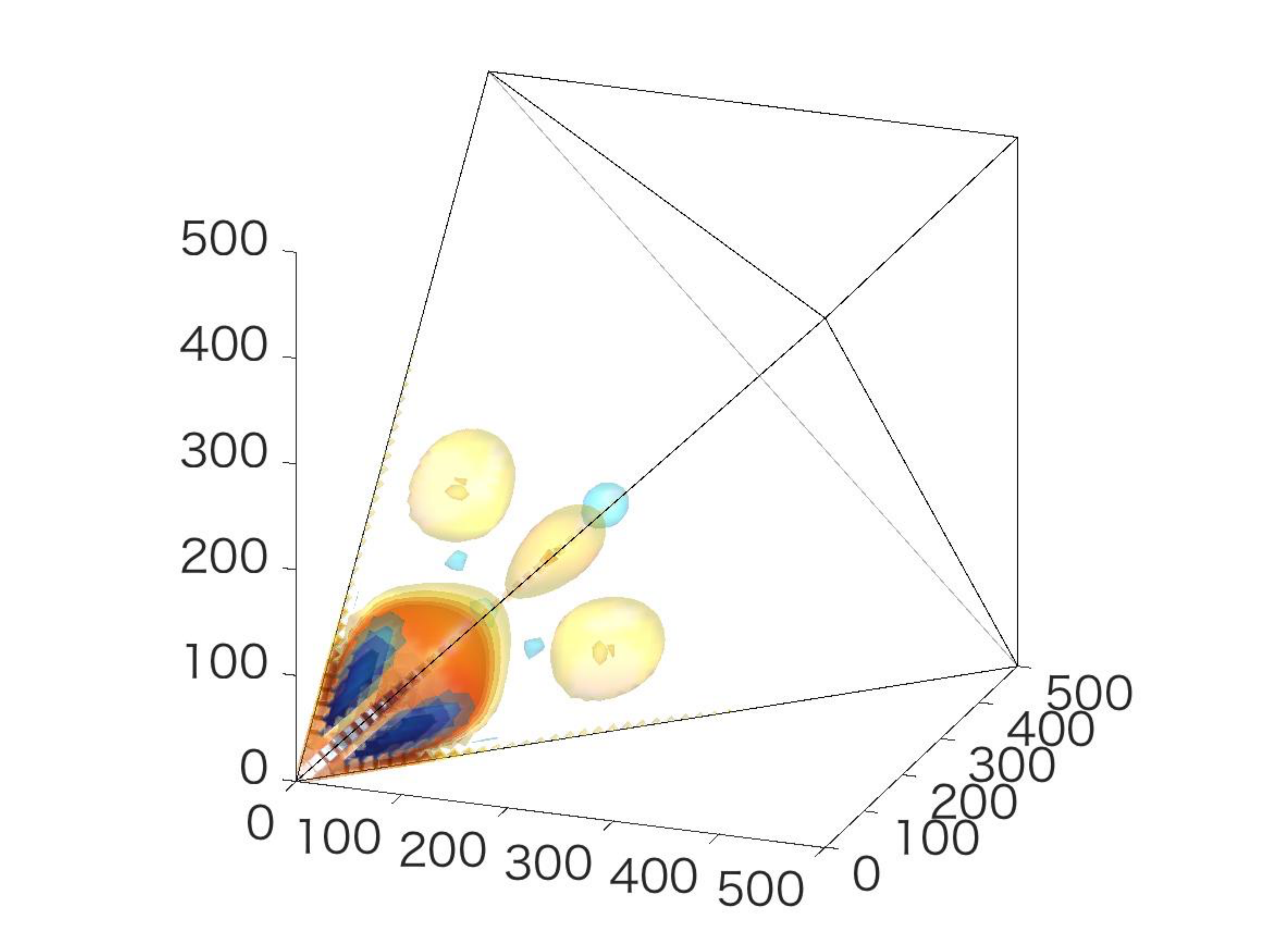}
  \end{center}
\end{minipage}
\end{tabular}
  \caption{Intensity distribution of the CMB temperature bispectrum in 3D $\ell$-space domain. Each axis corresponds to $\ell_1$, $\ell_2$, or $\ell_3$. Red (blue) dense regions correspond to large positive (negative) signal. Here, the bispectrum signal normalized by a constant Sachs-Wolfe bispectrum template \cite{Fergusson:2009nv} is plotted and hence dominant configurations are effectively highlighted.}\label{fig:3D} 
\end{figure}

Intensity distribution of the temperature bispectrum in 3D $\ell$-space domain is plotted in Fig.~\ref{fig:3D}. One can confirm there that the dominant signal comes from equilateral configurations $\ell_1 \sim \ell_2 \sim \ell_3$ (see Ref.~\cite{Shiraishi:2019yux} for illustrations of other shapes). However, it is highly suppressed for $\ell \gtrsim 100$ because of the cease of the ISW amplification. The other bispectra including E-mode and/or B-mode polarization, that are not described here, are also suppressed at small scales (because of the damping nature of the GW induced polarization field) and hence have similar intensity distributions.

\subsection{Detectability}

\begin{figure*}[t]
  \begin{tabular}{cc} 
   \begin{minipage}{0.5\hsize}
     \begin{center}
       \includegraphics[width=1\textwidth]{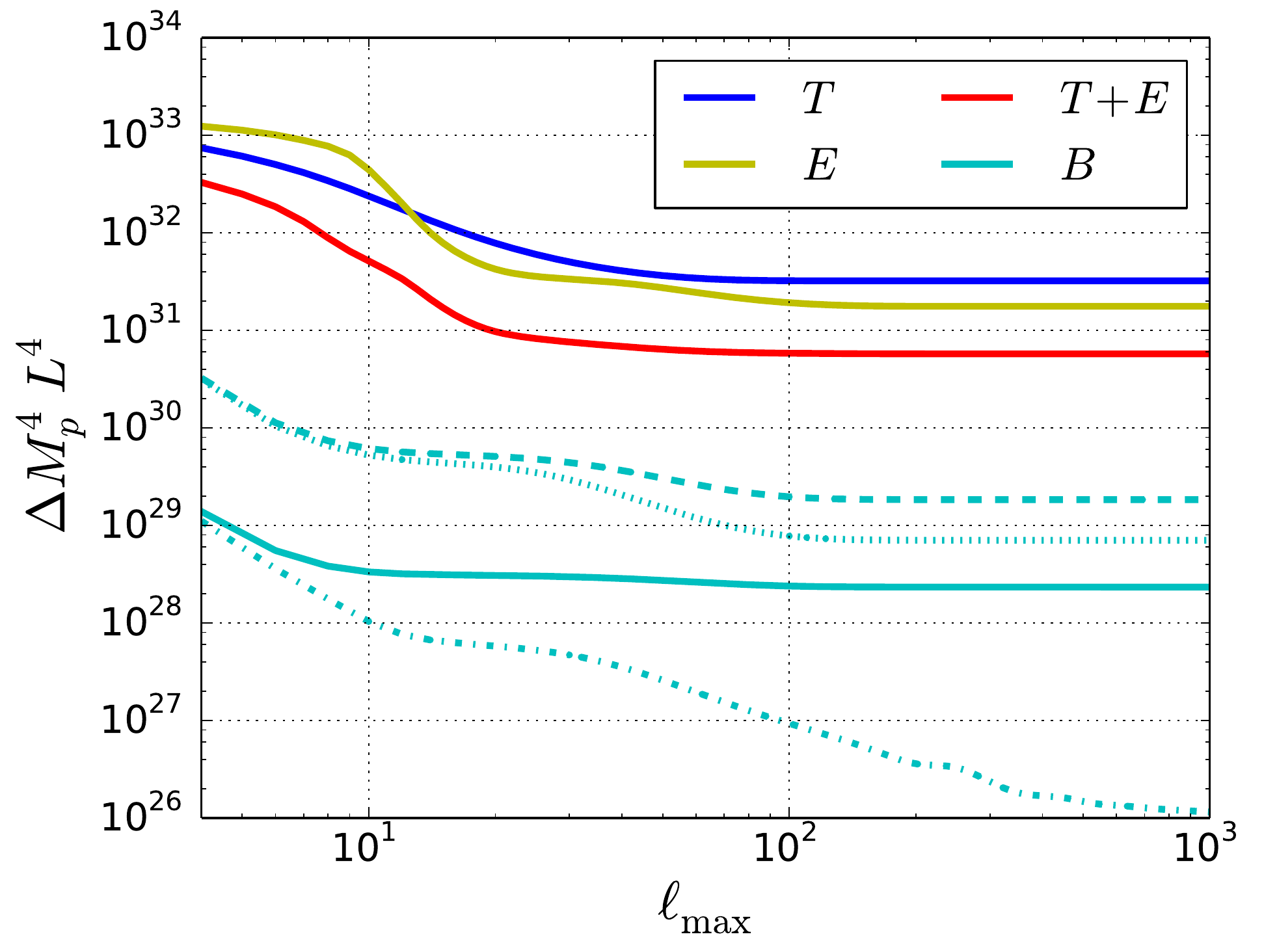}
  \end{center}
   \end{minipage}
   \begin{minipage}{0.5\hsize}
     \begin{center}
       \includegraphics[width=1\textwidth]{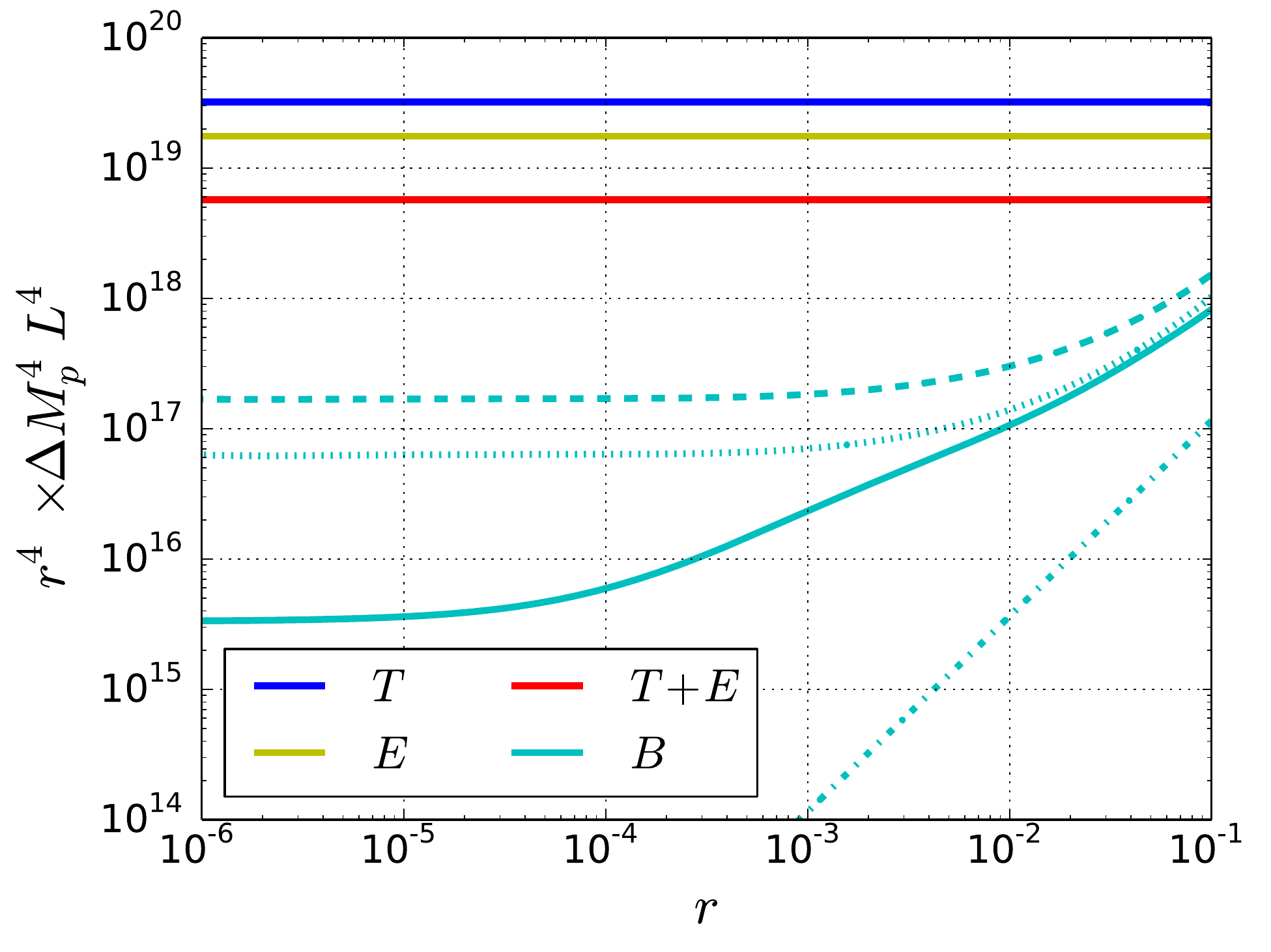}
  \end{center}
   \end{minipage}
  \end{tabular}
  \caption{Expected $1\sigma$ errors: $\Delta M_p^4 L^4$ from $T$-only (blue lines), $E$-only (yellow lines), $T+E$ (red lines) and $B$-only (cyan lines), as a function of the maximum multipole number $\ell_{\rm max}$ (left panel) and the tensor-to-scalar ratio $r$ (right panel). The results in the left and right panels are computed with $r = 10^{-3}$ and $\ell_{\rm max} = 1000$, respectively. Each line style of $B$-only (cyan line) discriminates the cleanliness level of the B-mode data: a perfectly-delensed and noiseless full-sky case (dot-dashed lines), a nondelensed and noiseless full-sky one (solid lines), a perfectly-delensed and LiteBIRD-like noisy one (dotted lines), and a nondelensed and LiteBIRD-like noisy one (dashed lines). For temperature and E-mode polarization, we assume a noiseless full-sky measurement.}
\label{fig:error}
\end{figure*}

In what follows, we focus on estimating a bispectrum amplitude parameter $M_p^4 L^4$ using the CMB bispectra~\eqref{eq:CMB_bis}. Then, assuming
\begin{equation}
  \left( \frac{H}{M_p} \right)^2 = \frac{\pi^2}{2} r A_S  \label{eq:h2}
\end{equation}
  with $A_S$ the scalar amplitude and $r$ the tensor-to-scalar ratio (see e.g. Ref.~\cite{lr}),  we rewrite $f_{k_1 k_2 k_3}$ into
\begin{equation}
f_{k_1 k_2 k_3} \equiv 360 \pi^8 A_S^4 r^4  M_p^4 L^4 (k_1 + k_2 + k_3)^{-6} . 
\end{equation}
We also fix $A_S$ to be an observed value, $2.2 \times 10^{-9}$ \cite{Aghanim:2018eyx,Akrami:2018odb}, and analyze the detectability of $M_p^4 L^4$ by varying $r$.

The Fisher matrices computed from auto bispectrum only ($T$-only, $E$-only or $B$-only) and temperature and E-mode polarization jointly ($T + E$) read, respectively,
\begin{eqnarray}
  F_X &=& \sum_{\ell_i}
  (-1)^{\ell_1 + \ell_2 + \ell_3}
  \frac{\left( \tilde{B}_{\ell_1 \ell_2 \ell_3}^{XXX} \right)^2 }{6 C_{\ell_1}^{XX} C_{\ell_2}^{XX} C_{\ell_3}^{XX}}  ~, \label{eq:fish_X}
  \\
  F_{T+E} &=& \frac{1}{6}
  \sum_{X_i, X_i' = T,E}
  \sum_{\ell_i} (-1)^{\ell_1 + \ell_2 + \ell_3}
  \tilde{B}_{\ell_1 \ell_2 \ell_3}^{X_1 X_2 X_3} (C_{\ell_1}^{-1})^{X_1 X_1'}
   \nonumber \\
  && \times  (C_{\ell_2}^{-1})^{X_2 X_2'} (C_{\ell_3}^{-1})^{X_3 X_3'} 
  \tilde{B}_{\ell_1 \ell_2 \ell_3}^{X_1' X_2' X_3'}  , \label{eq:fish_T+E}
 \end{eqnarray}
where $C_\ell^{X X'}$ is the CMB angular power spectrum, and $\tilde{B}_{\ell_1 \ell_2 \ell_3}^{X_1 X_2 X_3} = B_{\ell_1 \ell_2 \ell_3}^{X_1 X_2 X_3} / (M_p^4 L^4)$. Here, we have assumed that any NG contribution to the covariance matrix is negligibly small; thus, the covariance is diagonalized and given by the power spectrum alone. This would be justified as long as the Fisher matrices do not contain the information at very small angular scales. According to the {\it Planck} team~\cite{Akrami:2019izv}, the diagonal covariance matrix approximation works for $\ell \lesssim 2000$ by using {\it Planck} component separated temperature and polarization maps. For safety, our Fisher matrix computations are performed for $\ell \leq 1000$. The expected $1\sigma$ errors on $M_p^4 L^4$ from $T$-only, $E$-only, $T+E$, and $B$-only are given by $\Delta M_p^4 L^4|_T = 1/\sqrt{f_{\rm sky} F_T}$, $\Delta M_p^4 L^4|_E = 1/\sqrt{f_{\rm sky} F_E}$, $\Delta M_p^4 L^4|_{T+E} = 1/\sqrt{f_{\rm sky} F_{T+E}}$ and $\Delta M_p^4 L^4|_B = 1/\sqrt{f_{\rm sky} F_B}$, respectively, where $f_{\rm sky}$ is a sky coverage fraction. The B-mode power spectrum in $F_B$ is modeled by the sum of primordial contribution proportional to $r$, lensing B-mode one, and experimental noise spectrum, i.e., $C_\ell^{BB} = C_\ell^{\rm prim}(r) + C_\ell^{\rm lens}  + N_\ell$. Here, we consider four cleanliness levels of the B-mode data: a perfectly-delensed and noiseless full-sky case (i.e., $C_\ell^{\rm lens} = N_\ell = 0$ and $f_{\rm sky} = 1$), a nondelensed and noiseless full-sky one (i.e., $C_\ell^{\rm lens} \neq 0$, $N_\ell = 0$ and $f_{\rm sky} = 1$), a perfectly-delensed and noisy partial-sky one (i.e., $C_\ell^{\rm lens} = 0$, $N_\ell \neq 0$ and $f_{\rm sky} = 0.5$), and a nondelensed and noisy partial-sky one (i.e., $C_\ell^{\rm lens} \neq 0$, $N_\ell \neq 0$ and $f_{\rm sky} = 0.5$), are taken into account. Since an achievable level of delensing in the bispectrum analysis with future B-mode data has not been fixed, we simply examine two extreme cases: the perfectly-delensed and nondelensed ones. The noise spectrum is computed with experimental uncertainties due to beam, noise, mask and residual foreground expected in the LiteBIRD project \cite{Hazumi:2012aa,2016JLTP..tmp..169M,Matsumura:2013aja,Shiraishi:2016yun}.%
  \footnote{Comparable cleanliness level will also be achieved by other proposed CMB missions as CMB-S4 \cite{Abazajian:2016yjj} and CORE \cite{Delabrouille:2017rct}.}
  As was done in Ref.~\cite{Shiraishi:2016yun}, we assume that foregrounds arising from galactic dust emission and synchrotron radiation are subtracted by use of 9 channels planned in LiteBIRD (40-89\,GHz and 280-402\,GHz), achieving 2\% residual level in CMB maps. On the other hand, we assume a noiseless full-sky measurement of temperature and E-mode polarization and therefore $C_\ell^{TT}$, $C_\ell^{EE}$ and $C_\ell^{TE}$ do not contain the information of instrumental noise.

The left panel of Fig.~\ref{fig:error} depicts $\Delta M_p^4 L^4$ as a function of $\ell_{\rm max}$ at $r = 10^{-3}$. One can see there that $\Delta M_p^4 L^4|_{T}$, $\Delta M_p^4 L^4|_{E}$, $\Delta M_p^4 L^4|_{T+E}$, and $\Delta M_p^4 L^4|_{B}$ of the nondelensed case saturate at $\ell > 100$. This is because the tensor bispectrum (the numerator of $F$) rapidly decays for $\ell > 100$, where the covariance (the denominator of $F$) is still large due to the scalar-mode signal in $C_\ell^{TT}$, $C_\ell^{EE}$ and $C_\ell^{TE}$, and the lensing signal in $C_\ell^{BB}$. The similar behavior is seen in $\Delta M_p^4 L^4|_{B}$ of the perfectly-delensed and noisy case because of the dominance of $N_\ell$ for $\ell \gtrsim 100$. On the other hand, $\Delta M_p^4 L^4|_{B}$ of the perfectly-delensed and noiseless case can shrink as $\ell_{\rm max}$ increases because of the absence of the lensing signal and noise in $C_\ell^{BB}$.

The right panel of Fig.~\ref{fig:error} presents $\Delta M_p^4 L^4$ as a function of $r$ at $\ell_{\rm max} = 1000$. In the perfectly-delensed and noiseless case, $\Delta M_p^4 L^4|_B \propto r^{-5/2}$ holds for any $r$ since $\tilde{B}_{\ell_1 \ell_2 \ell_3} \propto r^4$ and $C_\ell^{BB} = C_\ell^{\rm prim}(r) \propto r$. Regarding the other error bars, since $C_\ell^{TT} \propto r^0$, $C_\ell^{EE} \propto r^0$ and
\begin{equation}
C_\ell^{BB} \simeq \begin{cases}
    C_\ell^{\rm prim}(r) \propto r & \text{for large $r$}\\
    C_\ell^{\rm lens} + N_\ell & \text{for small $r$} 
  \end{cases},
\end{equation}
the $r$ dependences are expected to be  
\begin{eqnarray}
  \Delta M_p^4 L^4|_{T} &\propto& r^{-4} ,  \\
  \Delta M_p^4 L^4|_{E} &\propto& r^{-4},  \\
  \Delta M_p^4 L^4|_{T+E}  &\propto& r^{-4}, \\  
  \Delta M_p^4 L^4|_B &\propto& 
  \begin{cases}
    r^{-5/2} & \text{for large $r$}  \\
    r^{-4}  & \text{for small $r$} 
  \end{cases} . 
\end{eqnarray}
These behaviors can be confirmed from each line in the right panel. We find there that, at any level of delensing, the graviton NG with 
\begin{equation}
M_p^4 L^4 \sim 10^{17} r^{-4} \label{eq:future}	
\end{equation}
is detectable by next decadal B-mode survey aiming at hunting $r < 10^{-2}$ such as LiteBIRD, CMB-S4 and CORE, and a smaller signal could be captured further in the future.

\subsection{Observational constraints}

Finally, we test this model with existing CMB data. We then follow a bispectrum estimation pipeline adopted in previous works on constraining other tensor NG shapes \cite{Shiraishi:2013wua,Shiraishi:2014ila,Ade:2015ava,Shiraishi:2017yrq,Akrami:2019izv}.

Here, $M_p^4 L^4$ is constrained from a coadded temperature data using WMAP 9-years V and W bands \cite{Bennett:2012zja,Hinshaw:2012aka}. The {\it Planck} data is not adopted for this analysis, while we expect from previous analyses of a very similar equilateral NG shape \cite{Shiraishi:2014ila,Ade:2015ava,Akrami:2019izv} that almost an identical limit is obtained from the temperature data, and it is slightly tightened by adding E-mode polarization data. We then employ an estimator \cite{Komatsu:2003iq} 
\begin{equation}
  \widehat{M_p^4 L^4} = \frac{1}{F_T}
  \sum_{\ell_i} 
  (-1)^{\ell_1 + \ell_2 + \ell_3}  
 \frac{\tilde{B}_{\ell_1 \ell_2 \ell_3}^{TTT} {\cal B}_{\ell_1 \ell_2 \ell_3}^{TTT} }{6 C_{\ell_1}^{TT} C_{\ell_2}^{TT} C_{\ell_3}^{TT}} ,  \label{eq:estimator}
\end{equation}
where 
\begin{eqnarray}
  {\cal B}_{\ell_1 \ell_2 \ell_3}^{TTT}
  &\equiv& \sum_{m_i} \left(
\begin{array}{ccc}
\ell_1 & \ell_2 & \ell_3 \\
m_1 & m_2 & m_3
\end{array}
\right)  
\left[
a_{\ell_1 m_1}^T a_{\ell_2 m_2}^T a_{\ell_3 m_3}^T
\right. \nonumber \\ 
&& \left. - 
\left( \Braket{a_{\ell_1 m_1}^T a_{\ell_2 m_2}^T}_{\rm MC} a_{\ell_3 m_3}^T + 2~{\rm perm.} \right)
\right]
\end{eqnarray}
is the bispectrum reconstructed from maps, and $\Braket{a_{\ell_1 m_1}^T a_{\ell_2 m_2}^T}_{\rm MC}$ denotes the variance of simulated Gaussian maps. This estimator properly works when the covariance matrix is approximately diagonalized. Such a situation is achieved by a recursive inpainting prefiltering technique \cite{Gruetjen:2015sta,Bucher:2015ura}, which has been implemented in map making for the {\it Planck} bispectrum estimation in order to avoid a time-consuming full-inverse covariance weighting process \cite{Ade:2015ava,Akrami:2019izv}. At the price of minimal ($\sim 5\%$) loss of optimality, we adopt this. Note that estimates from Eq.~\eqref{eq:estimator} rely on $r$, which however is singled out because of $\widehat{M_p^4 L^4} \propto r^{-4}$.

The best-fit value and the $1\sigma$ errors on $M_p^4 L^4$ are obtained computing Eq.~\eqref{eq:estimator} with observed data and 500 Gaussian simulations, respectively. Experimental features, i.e., beam shape and anisotropic noise, are added to the simulations by means of the WMAP-team methodology \cite{Komatsu:2008hk}. Moreover, both of the observed data and the simulations are masked using the KQ75 mask, leading to $f_{\rm sky} = 0.688$. All masked regions are inpainted after the removal of monopole and dipole components. The observed maps and the experimental tools are downloaded from the Lambda website \cite{Lambda}.

The above estimation process, unfortunately, comes at an unfeasibly high computational cost since a map-by-map computation of Eq.~\eqref{eq:estimator} already requires ${\cal O}(\ell_{\rm max}^5)$ numerical operations. This issue is actually solved if the summation in the 3D $\ell$-space domain is factorized as $\sum_{\ell_1 \ell_2 \ell_3} A_{\ell_1 \ell_2 \ell_3} = [\sum_{\ell_1} a(\ell_1)][\sum_{\ell_2} b(\ell_2)] [\sum_{\ell_3} c(\ell_3)]$. In fact, in the usual scalar-mode bispectrum analysis, the estimator sum is reduced straightforwardly since $\tilde{B}_{\ell_1 \ell_2 \ell_3}^{TTT}$ is given by a (partially) separable form \cite{Komatsu:2003iq}. In general situations where $\tilde{B}_{\ell_1 \ell_2 \ell_3}^{TTT}$ has a nonseparable structure (see Refs.~\cite{Shiraishi:2010kd,Shiraishi:2011st,Shiraishi:2012sn,Shiraishi:2013vha,Shiraishi:2013kxa,Shiraishi:2016yun,Meerburg:2016ecv,Domenech:2017kno,Tahara:2017wud,Bartolo:2018elp,Kothari:2019yyw} for tensor-mode NG cases), the modal approach \cite{Fergusson:2009nv,Fergusson:2010dm,Shiraishi:2014roa,Fergusson:2014gea,Shiraishi:2019exr} gives a solution. There, a nonseparable input bispectrum template is replaced by an approximated separable one generated via a modal decomposition based on finite numbers of separable eigenfunctions. This yields a fast estimator form imposing ${\cal O}(\ell_{\rm max}^3)$ numerical operations. In this paper, the modal approach is adopted because the input bispectrum under examination~\eqref{eq:CMB_bis} is obviously nonseparable. We perform the modal decomposition using 450 modal basis templates based on polynomials, and a $99.9\%$ reproduction of the input bispectrum is achieved.

Prior to working with the real data, we test the validity of our bispectrum estimation pipeline using simulations. We then produce ${\cal O}(10^2)$ NG maps including nonzero $M_p^4 L^4$ by means of the modal approach \cite{Fergusson:2009nv,Fergusson:2010dm,Shiraishi:2014roa}. Implementing all WMAP experimental features mentioned above in them, we do a map-by-map computation of Eq.~\eqref{eq:estimator}, and confirm that the average of output values recovers the input $M_p^4 L^4$. In a similar manner, we also compute Eq.~\eqref{eq:estimator} from ${\cal O}(10^2)$ Gaussian simulations, and confirm that the standard deviation of output values saturates a Cram\'er-Rao bound, $1 / \sqrt{f_{\rm sky} F_T}$ (see Fig.~\ref{fig:error} for values in a noiseless full-sky measurement). 

Since all validation tests are successfully passed as above stated, the real data are implemented in our pipeline. An analysis at a maximum angular resolution, $\ell_{\rm max} = 500$, yields 
\begin{equation}
  M_p^4 L^4 = (1.1 \pm 4.0) \times 10^{19} r^{-4} \ \ \
  (68\%\,{\rm CL}) . \label{eq:WMAP_limit}
\end{equation}
This comes from the foreground-cleaned data, while the raw data result in a comparable value. This result is unbiased by primordial scalar-mode bispectrum or secondary one (i.e., ISW-lensing bispectrum \cite{Hu:2000ee,Lewis:2011fk}) because it has a shape completely different from our template~\eqref{eq:CMB_bis}. These make our result robust. 


\subsection{Physical interpretations}

The prediction of the Vasiliev HS theory for the parameter $L$ is $L \simeq 0.4/H$, as reported in Eq.~\eqref{eq:hl}, from which it is clear that the signal is enhanced for a smaller value of the Hubble rate $H$. We can also use the relation between the tensor-to-scalar ratio and $H$ in Eq.~\eqref{eq:h2} as 
\begin{equation}
r \simeq 9 \times 10^7  \left( \frac{H}{M_p} \right)^2
\end{equation}
to rewrite the condition allowing the signal to be above the upper bound in Eq.~\eqref{eq:WMAP_limit} as
\begin{equation}
  H \gtrsim 2 \times 10^{-3} M_p \simeq 5 \times 10^{15} \, {\rm GeV},
\end{equation}
which is already ruled out by current constraints on the Hubble rate  $H< 6\times 10^{13} \, {\rm GeV}$ \cite{planck}. 

With future experiments, using Eq.~\eqref{eq:future}, the observability limit will be set to 
\begin{equation}
	H \gtrsim 5 \times 10^{-4} M_p \simeq 1 \times 10^{15} \, {\rm GeV}.
\end{equation}
In light of this, one can conclude that no observational confirmation of the Vasiliev HS gravity can be expected within the next generation of future experiments. This difficulty in observing such a signature in the specific case of the Vasiliev higher spin gravity theory is due to its specific prediction of the combined product $HL\simeq 0.4$. One could in principle hope that inflation took place with a lower Hubble scale $H$, enhancing the effect of the (Weyl)$^3$ interaction onto the graviton NG, but that, in turn, would decrease the amount of tensor modes reaching us (thus the dependence of the expected $1\sigma$ error on $r$) making any observation out of reach even for the planned future experiments.

However, a different theory of gravity predicting a different contribution from the (Weyl)$^3$ term to the three point function would turn out to be testable, provided that the length scale $L$ is predicted to be at least
\begin{equation}
	H L \gtrsim 8 ,
\end{equation}
where again we utilized the future detectability bound shown in Eq.~\eqref{eq:future}. Given the present bound on the Hubble rate \cite{planck}, then this condition translates into $L \gtrsim 3.2 \times 10^5 M_p^{-1}$. The nonobservation of such a signature would, on the contrary, put an upper bound on the scale $L$ as $L \lesssim 8/H$.

\section{Conclusions}\label{sec:con}

We have investigated the capability of measured CMB spectra to provide information on the strength of the  (Weyl)$^3$ term through the tensor NGs generated during inflation. Such term
is predicted to be relevant for the three point graviton correlator since it provides a conformal invariant tensor spectrum, together with the standard GR action. 

If the strength of the  (Weyl)$^3$ term is fixed by a length scale $L$, we have concluded that, with the currently available observations,  an upper limit is set at $M_p^4 L^4 = (1.1 \pm 4.0) \times 10^{19} r^{-4}$~(68\%\,CL), while the new generation of CMB experiments will be able to detect the presence of the  (Weyl)$^3$ term if $M_p^4 L^4 \gtrsim 10^{17} r^{-4}$, or $HL\gtrsim 8$. Unfortunately the well theoretically motivated higher spin theory \`a la Vasiliev does not satisfy such a bound and therefore will not be testable through the tensors NGs in the forthcoming observations.


\acknowledgements

V.\,D.\,L., G.\,F.  and A.\,R. are supported by the Swiss National Science Foundation (SNSF), project {\sl The Non-Gaussian Universe and Cosmological Symmetries}, project number: 200020-178787. M.\,S. is supported by JSPS Grant-in-Aid for Research Activity Start-up Grant Number 17H07319 and JSPS Grant-in-Aid for Early-Career Scientists Grant Number 19K14718. M.\,S. also acknowledges the Center for Computational Astrophysics, National Astronomical Observatory of Japan, for providing the computing resources of Cray XC50.

\bibliography{paper}

\end{document}